\documentclass[preprint]{aastex}	

\lefthead{Sakamoto, Fukuda, Wada, & Habe}
\righthead{Millimetric observations of M81}

\newcommand{\etal}{\mbox{\it et al.}}
\newcommand{\Msol}{\mbox{$M_\odot$}}
\newcommand{\Msolpc}{\mbox{$M_\odot$ pc$^{-2}$}}

\newcommand{\Lsol}{\mbox{$L_\odot$}}
\newcommand{\kms}{\mbox{km s$^{-1}$}}
\newcommand{\kmskpc}{\mbox{km s$^{-1}$ kpc$^{-1}$}}
\newcommand{\sqcm}{\mbox{cm$^{-2}$}}

\newcommand{\beam}{\mbox{beam$^{-1}$}}
\newcommand{\h}{\mbox{$^{\rm h}$}}
\newcommand{\m}{\mbox{$^{\rm m}$}}

\newcommand{\HH}{\mbox{H$_2$}}
\newcommand{\Halpha}{\mbox{H$\alpha$}}
\newcommand{\tm}[1]{\tablenotemark{#1}}
\newcommand{\nd}{\nodata}
\newcommand{\nuc}{\mbox{M81$^{*}$}}
\newcommand{\calib}{\mbox{0836+710}}
\newcommand{\rk}{\mbox{$R_{\rm nt}$}}
\newcommand{\maxrnilr}{\mbox{$\max(R_{\rm NILR})$}}

\newcommand{\Omegabar}{\mbox{$\Omega_{\rm bar}$}}

\begin{document}

\title{\small To appear in the Sept. 2001 issue of The Astronomical Journal}

\title{Millimetric Observations of the Center of M81 : \\
       A Starved Nucleus with Intraday Variability}

\author{
Kazushi Sakamoto\altaffilmark{1,2}, 
Hiroyuki Fukuda\altaffilmark{3},
Keiichi Wada\altaffilmark{3},
and
Asao Habe\altaffilmark{4}
}

\altaffiltext{1}{Nobeyama Radio Observatory, Minamisaku, Nagano, 
384-1305, Japan}
\altaffiltext{2}{Harvard-Smithsonian Center for Astrophysics, 
SubMillimeter Array, P. O. Box 824, Hilo, HI 96721, 
E-mail: ksakamoto@cfa.harvard.edu}
\altaffiltext{3}{National Astronomical Observatory, Mitaka, Tokyo, 181-8588, Japan}
\altaffiltext{4}{Department of Physics, Hokkaido University, Sapporo, 
060-0810, Japan}

\begin{abstract}
	The central kiloparsec of \objectname{M81} has been observed 
in the CO(J=1--0) line and the 3 mm continuum at 100 pc resolution 
in an attempt to probe molecular gas, 
and to search for the nuclear inner Lindblad resonance (NILR), 
around the low-luminosity AGN \nuc.
We found the following. 
(1) Molecular gas in the central kpc is mainly on a ``pseudoring'' or 
a spiral arm at a radius of about 500 pc.
(2) The region within  $\sim$300 pc from the nucleus is 
mostly devoid of molecular gas except for diffuse one; 
in particular, there is neither a giant molecular cloud that is now accreting 
on the nucleus nor a conspicuous gas feature 
that can be identified as an NILR.
(3) The 3 mm continuum emission shows significant intraday variation, 
suggesting an emitting region of $\sim 100$ AU.
(4) The $3\sigma$ upper limit for CO absorption toward the continuum source is 
$\int \tau_{\rm CO(0\rightarrow1)}\,dV < 0.1$ for a linewidth of 10 \kms.
The dearth of accreting molecular gas in the vicinity of the nucleus may 
explain the low luminosity of \nuc.
\end{abstract}


\section{Introduction \label{s.intro}}
M81 is a large ($L_{\rm B}= 2\times10^{10}$ \Lsol; Tully 1988), 
early-type (SA(s)ab; de Vaucouleurs \etal\ 1991), 
spiral galaxy at a distance of 3.6 Mpc \citep{Freedman94}.
At the center of M81 is an AGN called \nuc, for which an accreting massive 
black hole has been indicated by, among other things, 
a power-law X-ray continuum, a broad \Halpha\ line, 
a compact core-jet structure in radio wavelengths, 
and rapid variability of these features 
\citep[and references therein]{Ishisaki96,Bartel00,Bower96}.
In fact, M81 is one of the nearest galaxies that host an AGN 
with such strong evidence.
The luminosity of the nucleus, 
$L_{\rm 2-10 keV} \sim 2\times10^{40}$ erg s$^{-1}$,
$L_{\rm \Halpha} \sim 4\times10^{37}$ erg s$^{-1}$, and
$L_{\rm 6cm} \sim 1\times10^{20}$ W Hz$^{-1}$ str$^{-1}$,
places it among the low-luminosity AGN \citep{Ishisaki96,HFSp3,deBruyn76}.

An interesting issue that can be addressed from observations of molecular gas 
around \nuc\ is fueling to the active nucleus; 
e.g., whether the low-luminosity AGN is suffering from a shortage of fuel.
Also related to this issue is the search for the 
nuclear inner Lindblad resonance (NILR)\footnote{
The name of the resonance was ``nuclear Lindblad resonance (NLR)''
when it was first proposed. However, we rename it in this paper
to avoid confusion with the ``narrow line region'' whose
widely-used acronym in AGN studies is also NLR. 
The new name also makes it clear that the resonance is 
an inner Lindblad resonance (see \S 4).} 
that is predicted to form around a central massive black hole 
and help fuel it 
(Fukuda, Wada, \& Habe 1998; Fukuda, Habe, \& Wada 2000,
hereafter referred to as FWH98 and FHW00, respectively).
The proximity of M81 and the estimated mass of the central black hole, 
ranging from (3--8) $\times\; 10^{5}$ \Msol\ to $> 4\times 10^{7}$ \Msol\ 
\citep[and references therein]{Filippenko88, Iyomoto01} 
provide a good chance to detect the NILR.
Conveniently, M81 has a small nuclear bar of $\sim$1 kpc length 
that could drive the resonance \citep{Elmegreen95, Reichen94}.
In addition, millimeter continuum emission from \nuc, 
which is strong thanks to the galaxy's proximity, 
allows us to probe the tenuous molecular gas in front of the nucleus 
through absorption studies.
Further, monitoring the millimeter continuum during observations 
tells us the variability of \nuc.

CO emission from cold molecular gas in the center of M81 was first detected 
by \citet{Sage91} using the NRAO 12 m telescope.
No further observation has been reported, and, hence, the distribution of 
molecular gas around the nucleus has not been known to better than 1 kpc resolution.
Millimeter continuum emission from \nuc\ was detected once by \citet{Reuter96} 
but no monitoring has been reported.

We observed the central kpc (1\arcmin) of M81 at $\sim$100 pc resolution 
in CO(J=1--0) line and 3 mm continuum using the Nobeyama Millimeter Array (NMA).
The data revealed, for the first time, the distribution of molecular gas 
 in the vicinity of the nucleus, 
served as a monitor of the continuum emission from \nuc\ 
with a time resolution of $\sim$30 min, 
and also allowed us to search for cold molecular gas 
in front of the nucleus through absorption.

\section{Observations and Data Reduction \label{s.obs}}
	We observed the CO(J=1--0) line and 3 mm continuum using the NMA 
from November 1999 to January 2000 and in April 2000.
The six 10 m telescopes of the array are equipped with SIS receivers 
of single polarization and have 1\arcmin\ beamsize in FWHM.
They were pointed toward \nuc\
at  $\alpha_{\rm B1950} = 9\h 51\m 27\fs3004 $ 
and $\delta_{\rm B1950} = +69\arcdeg 18\arcmin 08\farcs267 $ \citep{Ma98}.
The log of observations is in Table \ref{t.obslog}.
We obtained seven transits in the most compact D configuration and one 
in the sparse AB configuration.
Two sets of correlaters, UWBC and FX, were used simultaneously, 
with UWBC covering 512 MHz (1300 \kms\ for CO) with 256 channels 
and FX covering 32 MHz (80 \kms\ for CO) with 1024 channels.
This setup was used to observe the broad CO emission line 
($\Delta V \sim 400$ \kms)
and to search for a narrow absorption line of CO against the 
continuum emission from \nuc.
The CO line was observed in the upper sideband (USB) at 115 GHz, 
while continuum data were obtained from the lower sideband (LSB) 
at 103 GHz, where the atmospheric attenuation is less severe.
Data from both sidebands were recorded in every 8 sec; 
this high rate was employed to reduce coherence loss in the measurements 
of continuum flux density.
In each transit, M81 and a quasar B\calib, which is $6 \arcdeg$ away from M81, 
were observed in 20- or 30-min cycles, and either 3C273 or 3C279 
was observed for $\sim$30 min for passband calibration.
Atmospheric attenuation was corrected by the chopper-wheel method.
Flux scale was established by comparing quasars with Uranus. 
The flux density of the calibrator B\calib\ was measured a total of 
14 times during our observations. 
It was 1.4 Jy in the first session of observations 
from November through January 
and 1.6 Jy in the second session in April, being constant during each session, 
with about $\pm 10$ \% error.

	Maps of CO emission were made using standard
data-reduction techniques for radio interferometry, 
including gain and passband calibration 
as well as continuum subtraction in UVPROC2, 
and Fourier transform and deconvolution in AIPS.
Data from the AB configuration were not used to make maps of CO emission, 
because little emission was detected in long baselines.
Resulting CO maps have a resolution of $6\farcs9 \times 5\farcs8$ 
(P.A.$= -36\arcdeg$), which corresponds to $ 120 \times 100$ pc 
at the distance of the galaxy.
The maps in this paper are not corrected for the primary beam response, 
but all flux measurements have been corrected for it.
	
	Absorption of CO has been searched in both UWBC and FX data.
UWBC data were binned every 2 channels to obtain 4 MHz (10.4 \kms) resolution, 
while FX data were binned every 16 channels to obtain 0.5 MHz (1.3 \kms) resolution.
Flux density of the nucleus was measured in each channel of the data cubes 
that were made without continuum subtraction, 
and provided spectra of the central 100 pc of M81.
Linear baseline was subtracted from each spectrum, 
to cancel small differences 
in spectral index between \nuc\ and the passband calibrators.
The rms of each spectrum is 0.032 and 0.14 for UWBC and FX spectrum, 
respectively, in the unit where the flux density of \nuc\ is unity.

	Variations in \nuc\ were monitored using the LSB data.
The LSB data were first averaged over the 512 MHz band, 
and then averaged over the 20- or 30-min of time between  
gain calibrator observations. 
The latter averaging was made for visibility amplitudes 
to avoid coherence loss. 
Arithmetic averaging of amplitudes (so-called `scalar averaging') was used 
when the signal-to-noise ratio of each visibility was high enough ($ > 7$), 
and the maximum likelihood method was used when it was not, 
thus avoiding bias due to amplitude averaging.
The LSB data for \calib\ were reduced in the same way, 
and used to remove any variation in system and atmospheric conditions.
The use of \calib\ as a gain calibrator here, as well as in the mapping above, 
has a built-in assumption that it had a constant flux density 
and was unpolarized during each transit. 
We believe the assumption is sound as discussed below.

\section{Results \label{s.results}}
\subsection{CO emission \label{s.COemission}}
	Figures \ref{f.chan} and \ref{f.sqash} show CO channel maps 
and the integrated intensity map, respectively.
The CO emission is greater than $3 \sigma$ from $-187$ \kms\ to 146 \kms, 
with nondetection in two channels near the systemic velocity.
The velocity range, as well as the depression near the systemic velocity, 
is in good agreement with the previous single-dish observations.
It is evident in the maps that CO emission is seen {\it around} 
the galactic center, mostly at galactocentric radii of between 300 and 600 pc.
The gas distribution is lopsided, and little emission is seen 
to the west of the nucleus.
For simplicity, we call the overall structure a ``pseudoring'' 
at a galactocentric distance of about 500 pc, 
though our data do not rule out other descriptions for the structure, 
such as spiral arm(s).
In fact, a somewhat similar structure in \Halpha\ was called 
``nuclear spiral'' in \citet{Devereux95}.
The pseudoring has a peak molecular gas surface density 120 \Msolpc\ 
at our resolution, where no correction for inclination has been made, 
and the following Galactic CO-to-\HH\ conversion relation is assumed
\citep{Sco87, Solomon87, Strong88}:
\begin{equation}
	\left(
	  \frac{\Sigma_{\rm mol}}{\Msolpc}
	\right) 
	= 
	6.8 \times 10^{2} 
	\left( 
	  \frac{I_{\rm CO}}{\rm Jy\, \kms\ arcsec^{-2}}
	\right).
\end{equation}
The total mass of molecular gas in the pseudoring is $1\times 10^{7} \Msol$,
and the structure is apparently made of several gas clumps.
The peak brightness temperature of each clump is $\sim$0.3 K 
when averaged over our beam size of $\sim$100 pc 
and a velocity width of 21 \kms, 
comparable to those for giant molecular clouds (GMCs) and GMC complexes 
in the outer regions of M81 \citep{Taylor98, Brouillet98}. 
It is also notable that the nucleus is devoid of CO emission.
The closest components to the nucleus are the GMCs seen 
in the $-124, -104$, and +42 \kms\ maps 
at the deprojected galactocentric radius of $\sim 300$ pc.
Their masses are $M_{\rm mol} \sim 5\times 10^{5} \Msol$.

Figure \ref{f.COspec} is the CO spectrum observed with NMA, 
whose primary beam is 60\arcsec\ in FWHM.
The total CO flux is 50 Jy \kms, which is half the single-dish flux 
observed with the NRAO 12 m telescope in its 55\arcsec\ beam
\citep{Sage91}.
Thus, provided that calibrations are accurate in both observations, 
we recovered about half the total CO flux; the rest must be 
either too extended or too faint (or both) to be detected 
with our interferometric observations.
If the missed emission is uniformly distributed in our field of view 
and has a linewidth of 20 \kms\ or smaller at each position, 
then its intensity in our channel maps would be 28 mJy \beam, 
while the rms in the channel maps is 13 mJy \beam.
It thus seems more likely that the missing flux is mainly due to 
a smoothly distributed population of small molecular clouds 
in the central kiloparsec.
Widespread distribution of small clouds, in addition to larger GMCs, 
has been suggested also on a spiral arm of M81 \citep{Brouillet98}.

The overall rotation seen in the channel maps is largely consistent 
with what is expected from the moderately inclined ($i= 58\arcdeg$) galaxy 
whose receding major axis is at P.A. = 330\arcdeg\ \citep{Adler96}.
Detailed velocity field structure is hard to discern 
in the limited spatial coverage of the current data, 
but obvious noncircular motions are not seen.
The CO linewidth of $\sim$350 \kms\ is already attained at a radius of 
$\sim$30\arcsec, while the linewidth of HI peaks at about 420 \kms\ 
at a radius of $\sim$400\arcsec\ \citep{Adler96}.
If the inner CO and outer HI disks are coplanar and neither of them 
is affected by noncircular motions, then the rotation curve of M81 
reaches 80\% of its maximum already at a radius of 500 pc. 
If the above assumptions hold, then the rotation curve is steeper 
in the central regions than estimated from previous HI observations
(Fig. 7 of Adler \& Westphal 1996), 
but is consistent with ones observed in spiral galaxies 
of similar luminosity \citep{Sakamoto99a}.
This is not surprising considering the lower spatial resolution of the 
HI observations.
The Keplerian dynamical mass calculated from the CO linewidth is
$M_{\rm dyn}(r \leq {\rm 500\;pc}) = 1 \times 10^{10}$ \Msol.

The active nucleus \nuc\ appears to be very close to the dynamical center
of the galaxy; a conservative upper limit to the offset is 10\arcsec\ or 200 pc.
Thus \nuc\ is not significantly displaced from the dynamical center 
despite its small mass, 
which \citet{Ho96} estimated to be 0.7 -- 3 $\times\; 10^{6}$ \Msol\
comparable to that of a GMC complex.

\subsection{Continuum emission \label{s.cont}}
	Continuum emission is detected only from the nucleus, 
and its flux density varied during our observations.
Figure \ref{f.D5cont}\ shows amplitudes of \nuc\ and \calib\ 
during observations on 2000 January 1 and 3.
The upper panels show data before the gain calibration using \calib\ 
but after the chopper-wheel calibration.
The data taken in better weather on January 1 show 
that the amplitude of \calib\ was constant to $\lesssim$5 \% 
during the 6-hr observations, 
in which the parallactic angle of the source changed almost linearly 
by $\sim$100\arcdeg.
The amplitude should have varied sinusoidaly if the source was 
linearly polarized,
because we observed in linear polarization mode with position angle
rotating with the parallactic angle.
The constant amplitude suggests that \calib\ had a constant flux density 
and a negligible degree of linear polarization, 
unless \calib\ varied to compensate for polarization, which is unlikely.
We also did not see significant variation in the amplitude of \calib\ 
during observations made on other dates in stable weather conditions.

The lower panels in Fig. \ref{f.D5cont}\ show data after gain calibration, 
in which a gain curve is obtained by interpolating the amplitude of \calib\ 
with a Gaussian kernel that has $\sigma = 0.015 $ day and a cutoff 
at $\Delta t = 0.050 $ day.
The gain calibration is necessary and useful, especially for data taken 
in changing weather conditions, such as those on January 3.
The common variation seen in the two sources on that date must be 
mostly due to varying radio seeing and coherence loss, 
judging from the correlation between the amplitudes and 
the scatter in visibility phases.
After the gain calibration, the amplitude of \nuc\ shows 
almost linear variation on both dates.
The lack of a sinusoidal pattern suggests that the linear polarization 
of \nuc\ is not large enough to dominate the amplitude variation, 
and that the amplitude variation is mainly due to 
the variation of the total flux density of \nuc.
	
	Table \ref{t.m81flux}\ lists the flux density of \nuc\ 
and its rate of variation for each night.
The most important result is that \nuc\ varies 
in almost every track of duration 5--10 hrs by a few to several 10 \%.
In particular, \nuc\ appears to have had a flare on J.D.= 2451456, 
as shown in Fig. \ref{f.flare}.
During our observing period of 152 days, the ratio of the maximum 
to the minimum flux density is $\sim$7, 
and the mean flux density is 0.39 Jy, 
a factor of 2 larger than the only previous measurement 
(at 3.4 mm) made in 1993 by \citet{Reuter96}.
It is also notable that variability is almost linear in each day 
and that we do not see significant variation with timescales 
of a few hours or less, 
even though the resolution of our observations is $\sim$0.5 hrs.

To characterize the timescale of the variation, doubling time is calculated 
as the time in which \nuc\ would have varied by a factor of 2 
if it had held its linear trend in variation in each track.
We do this extrapolation because \nuc\ did not show two-fold variation 
in any single track.
The largest variation seen in a single track is the 45\% increase 
during the 0.24-day observations on January 1.
The doubling timescales defined this way are $\sim$1 day, 
as listed in Table \ref{t.m81flux}.
Also, flux density (without extrapolation) doubled in 1.2 days 
during our first 2 days of observations.
Thus \nuc\ has intraday variability at 3 mm.

The variation of radio emission from \nuc\ has long been known 
in centimeter wavelengths \citep{Crane76, deBruyn76, Ho99, Bietenholz00}, 
with the degree of variation being about a factor of 2 over years 
and being smaller in longer wavelengths.
Although 40\% variation in 4 days was observed in 1974 by \citet{Crane76}, 
recent extensive observations did not agree on the variation 
in short timescales ($\sim$ days).
While \citet{Ho99} reported 10--60 \% variation on the timescale of 
$\lesssim$1 day at 3.6 cm, \citet{Bietenholz00} did not see 
variations larger than their 5\% standard error in their 12 epochs 
of 12--18 hr observations at the same wavelength.
Our observations show that intraday variability does exist 
at a millimeter wavelength. 
The larger degree of variation 
(a factor of 7 in our observing period of 5 months) is in accordance 
with the trend that the degree of variation increases with frequency.

The sizescale of the millimeter-emitting region should be below 
about 1 light-day or a few 100 AU from the light-crossing time argument, 
if there is no relativistic beaming.
This size is consistent with what is extrapolated from VLBI observations 
in centimeter wavelengths, 
where the size is 700 AU in 22 GHz and changes with frequency 
as $\nu^{-0.8}$ between 2.3 and 22 GHz \citep{Bietenholz96}.
It is also comparable to the size of the broad-line region 
in this galaxy \citep{Filippenko88,Ho96}.
For a source 1 light-day or smaller, 
the brightness temperature needs to be greater than $10^{11}$ K at 3 mm 
in order to have a flux density of 1 Jy. 
The 3 mm emission from \nuc\ is obviously nonthermal, 
as is the case with the centimeter emission, 
and the high brightness temperature is comparable to those 
observed in the cores of radio galaxies.

\subsection{CO absorption \label{s.COabsorption}}
	Figure \ref{f.contspec} shows spectra of \nuc, 
at high- and low-velocity resolutions, in the frequencies 
where CO(0$\rightarrow$1) absorption would be caused by molecular gas 
in either M81 or the Galaxy or both.
No absorption is seen in the spectra.
The 3$\sigma$ upper limit of the CO optical depth is 0.1 and 0.5 
for linewidths of 10.4 \kms\ and 1.3 \kms, respectively.
	
	Hydrogen column density toward \nuc\ has been estimated 
from X-ray observations to be $N_{\rm H} \sim 1\times 10^{21}$ \sqcm\ 
\citep{Ishisaki96, Pellegrini00}.
If the absorbing material is mostly molecular gas 
with a CO abundance of [CO]/[\HH] $= 10^{-4}$, 
then the integrated optical depth of CO, 
$\int \tau_{\rm CO(0\rightarrow1)}\,dV$, 
would be 10 and 1 \kms\ for LTE temperatures of 10 and 30 K, respectively.
The CO absorption should have been detected in the case of lower temperature gas.
Possible reasons for the nondetection therefore include 
ionized or atomic absorbing material, low CO abundance, 
high temperature ($> 30$ K) of molecular gas, 
and a large velocity width of the absorbing gas.

\section{Search for the Nuclear Inner Lindblad Resonance}
	Nuclear inner Lindblad resonance is the resonance between perturbations 
from a stellar bar and epicyclic motion of gas clouds orbiting 
around the galactic center that hosts a central massive black hole (FWH98).
It is formed, like other inner Lindblad resonances, on the radius at which 
$\Omega - \kappa/2$ is equal to \Omegabar, 
where $\Omega$ is angular frequency of the orbiting clouds, 
$\kappa$ is the epicyclic frequency, and
\Omegabar\ is the pattern speed of the bar.
Figure \ref{f.rotcur} shows a model rotation curve and its $\Omega - \kappa/2$ 
in a potential with a central massive black hole. 
It is the upturn of the rotation curve toward the central black hole 
that causes the upturn of the $\Omega - \kappa/2$ curve 
and makes the NILR possible.
As seen below, 
{\it the dynamical effect of a central massive black hole is far-reaching}, 
being able to form the NILR even at a radius 
within which only $\sim$1 \% of the total mass is due to the black hole.

\subsection{Maximum Radius of NILR}
	The maximum radius at which an NILR can be formed is  
where $\Omega - \kappa/2$ is its minimum\footnote{This is also
the minimum radius at which the inner inner Lindblad resonance (IILR) can
form. The two resonances can degenerate at this radius.}. 
This radius, \maxrnilr, is expected to be closely related to 
the radius where the rotation curve hits its minimum. 
We call the latter radius nuclear turnover and denote it as \rk. 
The nuclear turnover, \rk, is observable and can be easily understood 
as the radius where the mass of the central black hole is comparable 
to the mass of the other components inside it.
For these reasons, \rk\ must be a good scale to measure \maxrnilr.
In fact, there are cases where the ratio of the two radii, 
$\maxrnilr/\rk$, 
can be related to the relative mass of the central black hole 
with respect to the characteristic mass within the ordinary turnover 
radius of the rotation curve.

The rotation curve of a galaxy with a central massive black hole can be written, 
when normalized, in the form
\begin{equation}
	v_{\rm total}(r) = \sqrt{v_{g}^{2}(r) + \frac{m}{r}},
\end{equation}
where $r$ and $v$ are normalized radius and velocity, respectively, 
$v_{g}(r)$ is the normalized rotation curve of the galaxy 
without the central black hole, and $m$ is the normalized mass of the black hole.
It is assumed that the galaxy is not strongly barred.
We normalize radius so that the turnover radius of the rotation curve $v_g(r)$ 
is at $r\simeq 1$ and normalize the velocity so that 
the flat (or peak) rotation velocity is $v \simeq 1$.
Consequently, the parameter $m$ is roughly the ratio of the black hole mass 
to the galaxy mass within the turnover radius 
and should satisfy $m \ll 1$ for most disk galaxies.

Let us consider the case where the rotation curve without the central 
black hole can be well approximated, 
from the center through the turnover radius, 
with the first few terms of the Taylor-expansion, 
\begin{equation}
	v_{g}(r) = ar + br^2 + cr^3 + \cdots.
\end{equation}
The coefficient $a$ is positive and of the order of unity, 
and the first non-zero coefficient after $a$ is negative 
and is also of the order of unity.
An example of this class is the rotation curve of the logarithmic potential,
\begin{eqnarray}
	v_{g,\log}(r) 
	& = & 
	\frac{r}{\sqrt{1 + r^2}} \\
        & = &
        r -\frac{1}{2}r^3 + \cdots,
\end{eqnarray}
where $a=1$, $b=0$, and $c=-1/2$.
The radius of nuclear turnover is, for small $m$,
\begin{equation}
	r_{\rm nt} \approx \left( \frac{m}{2a^{2}} \right)^{1/3}.
\end{equation}
The maximum radius of the nuclear inner Lindblad resonance is, 
from the minimum of the $\Omega - \kappa/2$ curve,
\begin{equation}
	\max(r_{\rm NILR}) \approx \left( -\frac{9m}{2ab} \right)^{1/4}
\end{equation}
if $b$ is not zero.
It is 
\begin{equation}
	\max(r_{\rm NILR}) \approx \left( -\frac{9m}{8ac} \right)^{1/5}
\end{equation}
if $b$ is zero but $c$ is not.
The ratio of the two radii is, therefore,
\begin{equation}
  \frac{\max(r_{\rm NILR})}{r_{\rm nt}}
  \approx
  \left\{   
  \begin{array}{ll}
     1.8 \; m^{-1/12}\; a^{5/12}\; (-b)^{-1/4} & \mbox{if $b\neq 0$}\\
     1.3 \; m^{-2/15}\; a^{7/15}\; (-c)^{-1/5} & \mbox{if $b=0$ and $c \neq 0$}.
  \end{array}
\right. 
\label{eq.exp.ratio}
\end{equation}
Ignoring the coefficients $a,b,c$, which are of the order of unity, 
the ratio is in the range of 2--8 for the fractional mass range of 
$m = 10^{-2}$ -- $10^{-6}$.
The weak dependence of $\max(r_{\rm NILR})$ on the black hole mass $m$ 
is notable, 
and the nearly constant value of the $\max(r_{\rm NILR})/r_{\rm nt}$ 
for the most realistic range of $m$ ensures 
that $r_{K}$ can be conveniently used to measure the radius of the NILR.

There are rotation curves for which Taylor expansion is not possible 
around $r=0$ and therefore the above analysis does not apply. 
However, some of those rotation curves do have similar characteristics on NILR 
as in the cases discussed above.
An example is the rotation curve of an exponential disk
with surface density $\Sigma(R) = \Sigma_{0}\exp (R/R_d)$,
\begin{equation}
	v_{g,\exp}(r) 
	 =  
	2^{3/2} r \left[ I_0(r)K_0(r) - I_1(r)K_1(r) \right]^{1/2},
\end{equation}
where the radius is normalized as $r \equiv R/2R_d$, and $I$ and $K$ are 
modified Bessel functions (see \citet{BT87}\ p.77). 
The turnover radius of the rotation curve is $r\approx 1.1$.
The conditions that approximately describe $r_{\rm nt}$ 
and $\max(r_{\rm NILR})$ are, 
for small $m$,
\begin{equation}
	r_{\rm nt,\;exp}^{3} \left( - \log r_{\rm nt,\;exp} \right)
	 \approx  
	\frac{m}{16}
\end{equation}
and 
\begin{equation}
	\left[\max(r_{\rm NILR,\;exp})\right]^{3} 
	\left( - \log r_{\rm NILR,\;exp} \right)^{-1}
	 \approx  
	\frac{9 m}{4},
\end{equation}
respectively.
They lead to the formula
\begin{equation}
	\frac{\max(r_{\rm NILR,\;exp})}{r_{\rm nt,\;exp}}
	 \approx 
	\{ 36 \log\left[\max(r_{\rm NILR,\;exp})\right] 
	       \log r_{\rm nt,\;exp} \}^{1/3}
\end{equation}
that shows the ratio being larger for smaller $m$ 
but being very weakly dependent on $m$, 
as is the case for eq. (\ref{eq.exp.ratio}).
Numerically, the ratio is in the range of 5--9 for $m = 10^{-2}$ -- $10^{-6}$.

The parameter $m$ for M81 is estimated to be 
$m \sim 5 \times 10^{-5}$ -- $5\times 10^{-3}$, 
by use of $M_{\rm dyn}(r \leq 500\; {\rm pc}) = 1 \times 10^{10}$ \Msol\ 
obtained from our CO observations and the black hole mass
in the literature (see \S \ref{s.intro}).
For this range of $m$, as we have seen, NILR can be formed 
at up to several times larger radius than the nuclear turnover.
The nuclear turnover radius, \rk, can be constrained as $\rk \leq $ 15 -- 70 pc 
by using the assumption that the mass density is uniform in the central kpc.
It is an upper limit because the density most likely increases 
toward the galactic center, as readily inferred from light distribution.
Combining these numbers, we can estimate \maxrnilr\ to be 100 pc 
for a smaller number for the black hole mass and 300 pc for a larger number.
Thus it is unlikely that the gas feature (pseudoring) at around 500 pc 
is due to an NILR.
Note that the radius of an NILR can be much smaller than the upper limits 
(and \rk) for a large \Omegabar.

\subsection{Numerical Simulations}	
	We made numerical simulations for gas around an NILR by using parameters 
adjusted for M81 in order to tell what can be seen if there is an NILR
in the galaxy.
The simulation is non-selfgravitating, isothermal, and two-dimensional.
Evolution of a rotating gas disk is simulated in a time-independent, 
external potential having a bar and a central black hole.
The potential is in the form 
\begin{equation}
  \Phi_{\rm ext} \equiv \Phi_{\rm BH} + \Phi_{\rm bar} + \Phi_{\rm disk},
\label{eq.potential}
\end{equation} 
where $\Phi_{\rm BH}$, $\Phi_{\rm bar}$, and $\Phi_{\rm disk}$ are 
a central massive black hole, a stellar bar, 
and a disk, respectively.
They are defined as follows:
\begin{equation}
  \Phi_{\rm BH} \left(R\right)   
  \equiv 
  -\frac{GM_{\rm BH}}{(R^2+ a^2)^{1/2}},
\end{equation}
\begin{equation}
  \Phi_{\rm disk} \left(R\right) 
  \equiv  
  - \frac{3^{3/2}}{2} \frac{a_{d} v_{0}^{2}}{(R^{2}+ a_{d}^{2})^{1/2}},
\end{equation}
and
\begin{equation}
  \Phi_{\rm bar} \left(R,\phi\right)  
  \equiv  
  - \frac{3^{3/2}}{2} \frac{a_{b} v_{0}^{2}}{(R^{2}+ a_{b}^{2})^{1/2}} 
  \left[1+ \varepsilon_0 \frac{a_b R^2}{(R^2+a_b^2)^{3/2}} \cos2\phi \right], 
\end{equation}
with $a=10$ pc, $a_d=5$ kpc, $a_b = 0.5$ kpc, 
$v_0= 200$ \kms, and $\varepsilon_0 = 0.05$.
This model is based on the potentials used in \citet{Wada98}.
Figure \ref{f.rotcur} shows the rotation curve and $\Omega - \kappa /2$ of 
the model potential with $M_{\rm BH} = 10^7$ \Msol. 
The initial gas disk is axisymmetric and rotationally supported, 
with a uniform surface density and total mass of $10^7$ \Msol\ 
within a 2 kpc diameter.
The isothermal equation of state is adopted 
along with a sound velocity of 10 \kms.

We use the second-order Euler mesh code, 
which is based on the Advection Upstream Splitting Method 
(AUSM; Liou \& Steffen 1993),  to solve the hydrodynamical equations.
Details of the numerical scheme are described in \citet{Wada01}.
Simulations are made on $512\times512$ Cartesian grid points 
in a region of $2 \times 2$ kpc; the spatial resolution is 3.9 pc.

Figure \ref{f.sim} shows time evolution of gas for $\Omegabar = 50$ \kmskpc\
and $M_{\rm BH} =10^6, 10^7$, and  $10^8 M_\odot$.
Table \ref{t.radii} lists \rk, $R_{\rm NILR}$, and $\max(R_{\rm NILR})$
for each mass of the black hole.
The small concentration of gas that develops near the galactic center 
at $R \sim R_{\rm NILR}$ is due to the NILR. 
The two-armed spiral structure is more clearly seen 
in cases with larger $M_{\rm BH}$. 
Evolution of gas around an NILR, especially 
when the self-gravity of gas is important, 
has been studied in more detail in FHW00 and in FWH98.
Outside NILR, the inner inner Lindblad resonance (IILR) creates 
an oval-shaped gas distribution, two leading spirals, 
with its major axis leading the bar major axis by about 45\degr.

\subsection{Comparison with Observations}
	The lack of CO emission around the nucleus, 
inside the 500 pc pseudoring, prevents us from detecting the features of an NILR.
Among the reasons that can possibly explain the absence of these features are  
the following: 
1) the NILR had quickly driven most of gas around the nucleus to the AGN 
as suggested in FWH00; 
2) the gas around the NILR is not in molecular form; 
3) the NILR is at a too small radius for us to detect the gas accumulated 
around the resonance; 
and 
4) the NILR does not exist because of a slow pattern speed 
or weakness of the bar perturbation.
Despite the nondetection in CO, HST images in optical revealed 
a spiral dust lane extending $\sim$12\arcsec\ (200 pc) 
to the north of the nucleus, 
and a disk-like feature of $\sim$7\arcsec\ (120 pc) diameter 
in \Halpha\ \citep{Devereux97,Pogge00}. 
These features could be related to an NILR. 
Kinematical information is needed to confirm the possibility.

The pseudoring (or spiral feature) in molecular gas 
at $R \sim 500$ pc is suggestive of a gas dynamical mechanism 
to arrange gas clouds --- possibly an inner Lindblad resonance.
The current data, however, are too scarce to pinpoint the mechanism, 
other than to say that it is unlikely to be an NILR.
The dynamics in the central kiloparsec may well be affected 
by the nuclear bar, or oval distortion, seen in the near-infrared, whose
semi-major axis is $\sim$30\arcsec\ (0.5 kpc) at P.A. = 142\arcdeg\
\citep{Elmegreen95}.
The observed gas distribution, however, 
does not show twin peaks and/or gas ridges at the bar's leading side 
unlike many barred galaxies.
The nuclear bar may have too small a quadrupole moment to create those features 
or may have a deviation from bisymmetry, which 
causes the lopsided gas distribution.

\section{Starved Nucleus \label{s.nuclues}}
	M81 does not have much molecular gas in the galactic center.
The gas mass in the central kiloparsec is $\sim 2\times 10^{7}$ \Msol, 
which is more than an order of magnitude smaller than the number 
in more gas-rich spiral galaxies \citep{Sakamoto99a,Sakamoto99b}.
The mass fraction of molecular gas in the central kiloparsec is 
$M_{\rm mol}/M_{\rm dyn}(r \leq 500\; {\rm pc}) \approx 2\times10^{-3}$, 
where missing flux is taken into account.
The low fraction is in accordance with the trend 
that galactic nuclei with an optical spectrum indicative of an AGN 
tend to have a lower gas mass fraction than do galactic nuclei 
with an HII type spectrum \citep{Sakamoto99b}.
The 3$\sigma$ upper limit for the mass of molecular gas 
in the central 100 pc is 
$M_{\rm mol}^{\rm nuc} \leq 4\times 10^{5} (\Delta V/ 100\,\kms)^{1/2} \Msol$, 
where $\Delta V$ is the CO linewidth.
Thus, it appears that \nuc\ is starving for infalling interstellar medium.

The lack of significant CO absorption toward \nuc\ is consistent 
with its type 1 nature in the framework of the unified scheme, 
in the sense that the putative accretion disk or broad line region 
does not suffer much obscuration from surrounding material.
On the other hand, the lack of CO emission at the nucleus prevented us 
from verifying if there is a gas torus or disk around \nuc.

With the paucity of fueling molecular gas around the active nucleus, 
its activity may well be limited by the intermittent supply of fuel.
If the AGN flares up when, for example, the closest GMC starts to 
interact with the nucleus, the timescale of the variability 
would depend on the interval between such interactions.
If a bar-streaming mechanism that radially transports 
the nearest GMC complexes to the nucleus is at work, 
as was observed in NGC 5005 \citep{Sakamoto00}, 
then the timescale would be of the order of Myr ($\approx$ 100 pc/100 \kms).
On the other hand, if dynamical friction to the GMCs is 
the only mechanism that transports the clouds, 
then \nuc\ is really starving and will not get the next large 
supply of fuel for $10^{9-10}$ yrs.

Finally, we note that a $5\times 10^5$ \Msol\ black hole, 
like GMCs of similar mass, takes $10^{9-10}$ yrs to fall from 
$R \sim 300$ pc to the dynamical center through dynamical friction.
Thus, if the mass of \nuc\ is close to the lower value in the literature, 
\nuc\ should be older than 1 Gyr or was formed at the very center of 
the galactic potential or both.
Tighter constraints from further observation of this nearby AGN 
would be useful in studies of the formation and evolution of AGNs.

\section{Summary \label{s.summary}}
	The central kiloparsec of the nearby spiral galaxy M81 
has been observed in CO(J=1--0) and, at the same time, 
its low-luminosity AGN \nuc\ has been monitored in the 3 mm continuum.
A search for NILR has also been carried out for M81.
Our observations and analysis yield the following results:
\begin{enumerate}
\item	Most of the molecular gas in the central kiloparsec is detected 
on a ``pseudoring'' or a nuclear spiral at the galactocentric radius of 
$\sim 500$ pc.
The area within 300 pc from \nuc, which appears to be at the dynamical center, 
is devoid of CO emission, except for diffuse emission, which can not be detected 
with our observations.
In particular, no CO emission is detected at the nucleus nor is any CO absorption 
detected against the continuum emission from \nuc.
The upper limit for the gas mass in the central 100 pc is 
$M_{\rm mol}^{\rm nuc} \leq 4\times 10^{5} (\Delta V/ 100\,\kms)^{1/2} \Msol$.
The GMCs nearest to the nucleus are 300 pc away, 
hence the nucleus will not be fueled by them at least in the next Myr.
	
\item	The 3 mm continuum from \nuc\ showed conspicuous variation; 
a factor of 7 in 5 months and up to 45\% in 6 hrs.
The intraday variation suggests $\lesssim$100 AU size 
and $\sim$10$^{11}$ K brightness temperature for the emitting region.
	
\item	It is pointed out that the maximum radius of the NILR 
can be several times larger than the radius 
at which the rotation curve turns over from Keplerian to rigid-body rotation.
Despite this factor, the pseudoring at $R \sim$500 pc is too far 
from the nucleus to be due to the NILR.
No CO feature that can be associated with an NILR is found in M81.
\end{enumerate}

\vspace{5mm}	

        We are grateful to NRO staff for their help during observations.
The NASA Extragalactic Database was used to carry out our research.


\clearpage	

\begin{deluxetable}{lccccccc}
\tablecaption{Log of observations \label{t.obslog}}
\tablewidth{0pt}
\tablehead{
 \colhead{Date} & 
 \multicolumn{2}{c}{Time} &
 \multicolumn{2}{c}{H.A.} &
 \colhead{$t_{\rm middle}$} & 
 \colhead{Config.} &
 \colhead{$t_{\rm cycle}$} 
\\
 \colhead{ } & 
 \colhead{Start} &
 \colhead{End} &
 \colhead{Start} &
 \colhead{End} &
 \colhead{(J.D.)} & 
 \colhead{ } &
 \colhead{min} 
}
\startdata
1999 Nov 29	& 18\h58\m	& 25\h 10\m & $-1\fh2$ & 5\fh0 & 2451512.42 & D  & 30 \\
1999 Nov 30    & 17 10 & 23 25 & $-2.9$ & 3.3 & 2451513.35 & D  & 30 \\
1999 Dec 27    & 17 24 & 23 25 & $-0.9$ & 5.1 & 2451540.35 & D  & 20 \\
2000 Jan \phn 1    & 17 37 & 23 25 & $-0.3$ & 5.4 & 2451545.35 & D  & 20 \\
2000 Jan \phn 3    & 16 41 & 23 25 & $-1.2$ & 5.6 & 2451547.34 & D  & 20 \\
2000 Jan 11    & 13 37 & 22 55 & $-3.7$ & 4.4 & 2451555.26 & AB & 20 \\
2000 Apr 18    & 06 45 & 16 25 & $-4.1$ & 5.5 & 2451652.98 & D  & 20 \\
2000 Apr 29	& 06 06	& 14 25	& $-4.1$ & 4.2 & 2451663.93 & D  & 20 \\
2000 Apr 30	& 05 28	& 14 25	& $-4.6$ & 4.3 & 2451664.91 & D  & 20   
\enddata
\tablecomments{Observing date and time are in UT.
Hour angle of M81 is in hours.
In each cycle of duration $t_{\rm cycle}$, gain calibrator is  
observed for 3 min; 2 min is used to slue telescopes; 
and M81 is observed for the rest. }
\end{deluxetable}

\clearpage	

\begin{deluxetable}{lcccc}
\tablecaption{Flux density of M81$^{*}$ \label{t.m81flux}}
\tablewidth{0pt}
\tablehead{
 \colhead{$t_0$} & 
 \colhead{$\Delta T_{\rm obs}$} &
 \colhead{$\overline{S_{\rm 3mm}}$} &
 \colhead{$(d S_{\rm 3mm}/dt)/ S_{\rm 3mm}(t_0)$ } &
 \colhead{$\tau_2$} 
\\
 \colhead{(JD)} & 
 \colhead{day} &
 \colhead{Jy} &
 \colhead{day$^{-1}$ } &
 \colhead{day}
}
\startdata
2451512.29    & 0.26 & 0.25 & $\phm{-} 0.62 \pm 0.18$ & 1.6 \\
2451513.22    & 0.26 & 0.40 & $\phm{-} 1.66 \pm 0.18$ & 0.6 \\
2451540.22    & 0.25 & 0.51 & $\phm{-} 0.22 \pm 0.18$ & 4.5 \\
2451545.23    & 0.24 & 0.63 & $\phm{-} 1.91 \pm 0.09$ & 0.5 \\
2451547.20    & 0.28 & 0.98 & $       -0.32 \pm 0.05$ & 1.6 \\
2451555.07    & 0.39 & 0.35 & $       -0.71 \pm 0.13$ & 0.7 \\
2451652.78    & 0.40 & 0.25 & $\phm{-} 0.56 \pm 0.14$ & 1.8 \\
2451663.75    & 0.35 & 0.15 & \nd              & \nd \\
2451664.73    & 0.37 & 0.21 & $\phm{-} 1.37 \pm 0.19$ & 0.7
\enddata
\tablecomments{Observations of M81 start at time $t_0$ and last for
the duration of $\Delta T_{\rm obs}$.
Mean flux density of \nuc\ during observations is $\overline{S_{\rm 3mm}}$,
with the absolute calibration error of about 10\%.
Flux densities in each transit are fitted with a linear function, 
$ S_{\rm 3mm}(t) = S_{\rm 3mm}(t_0) + \frac{dS_{\rm 3mm}}{dt} (t-t_0) $. 
Gain correction has been applied before fitting.
No fitting was made for the data taken on J.D. = 2451663, because of poor
signal-to-noise ratio. 
The timescale $\tau_2$ is the time in which the flux density would double 
or become half if the linear variation held.}
\end{deluxetable}

\clearpage	

\begin{deluxetable}{ccccc}
\tablecaption{Key radii in the model \label{t.radii}}
\tablewidth{0pt}
\tablehead{
 \colhead{$M_{\rm BH}$} & 
 \colhead{\rk} & 
 \colhead{$R_{\rm NILR}$} &
 \colhead{$\max(R_{\rm NILR})$} &
 \colhead{$R_{\rm IILR}$}
\\
 \colhead{\Msol} & 
 \colhead{pc} &
 \colhead{pc} &
 \colhead{pc} &
 \colhead{pc} 
}
\startdata
$10^{6}$  & 17	& 28 				& \phn84 & 279\\
$10^{7}$  & 37 	& 82 				& 137 	 & 284\\
$10^{8}$  & 82	& \nodata\tm{\dag} 	& 237	 & \nodata 
\enddata
\tablecomments{Key radii in the simulations shown
in Fig. \ref{f.sim}. The model potential
is eq. (\ref{eq.potential}) and the pattern speed of 
the bar is $\Omegabar = 50$ \kmskpc. 
See text for other parameters in the model.}
\tablenotetext{\dag}{The minimum of $\Omega - \kappa/2$ is 
60 \kmskpc\ and hence, strictly speaking, an NILR does not exist.
However the characteristic gas feature of NILR appears at
radii where $\Omega - \kappa/2 \approx \Omegabar$.}
\end{deluxetable}


\clearpage
\setcounter{figure}{0}

\begin{figure}[t]
{\hfill
\epsscale{0.8}                 
\plotone{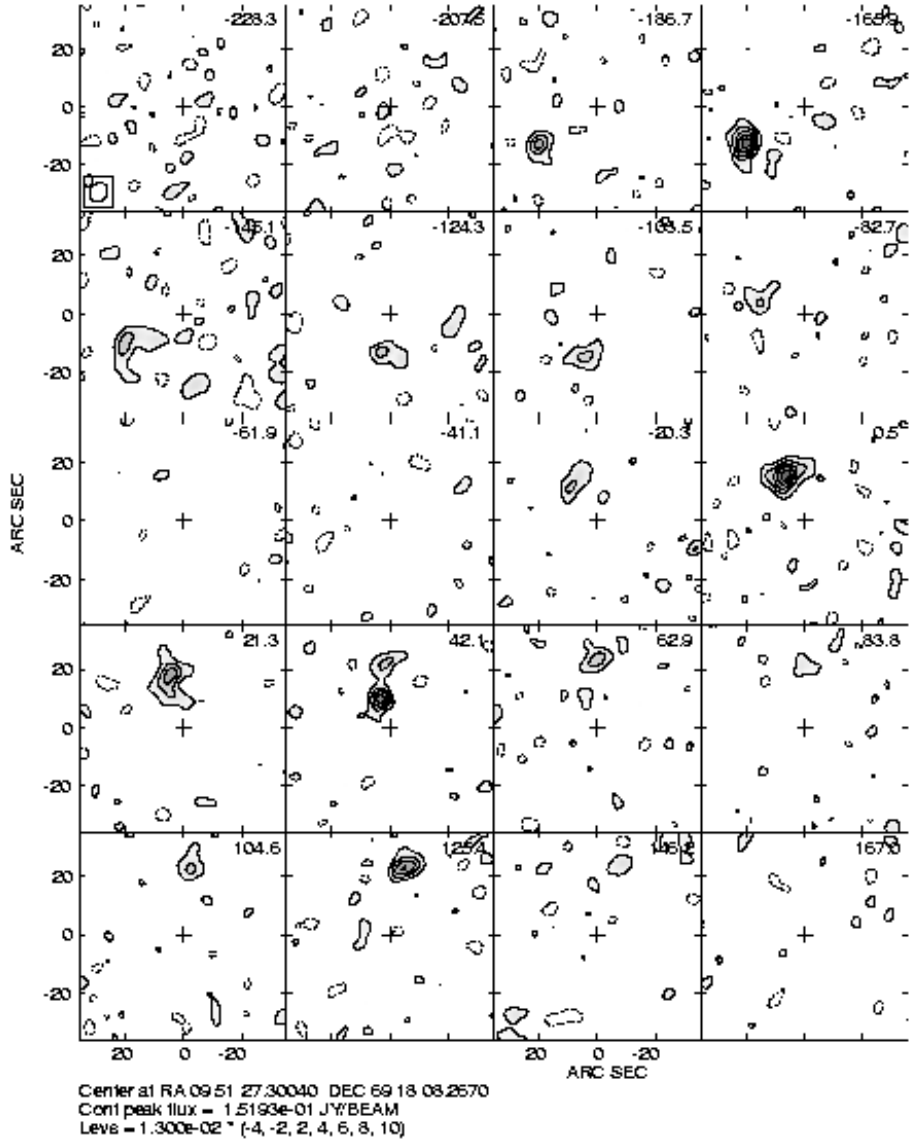}       
\hfill}
\caption{
CO(J=1--0) channel maps of 20.8 \kms\ resolution in the central kpc of M81. 
Continuum emission from the nucleus (shown as a cross) has been subtracted. 
Contours are at $-4,-2,2,4,6,8$, and 10 $\times$ 13 mJy \beam\ 
($= 30$ mK $= 1\sigma$).
LSR velocity (\kms) is shown in each panel.
The synthesized beam of $6\farcs9 \times 5\farcs8$ in FWHM 
is shown in the bottom left corner of the first panel. 
\label{f.chan}}
\end{figure}

\clearpage

\begin{figure}[t]
{\hfill 
\epsscale{0.7}                
\plotone{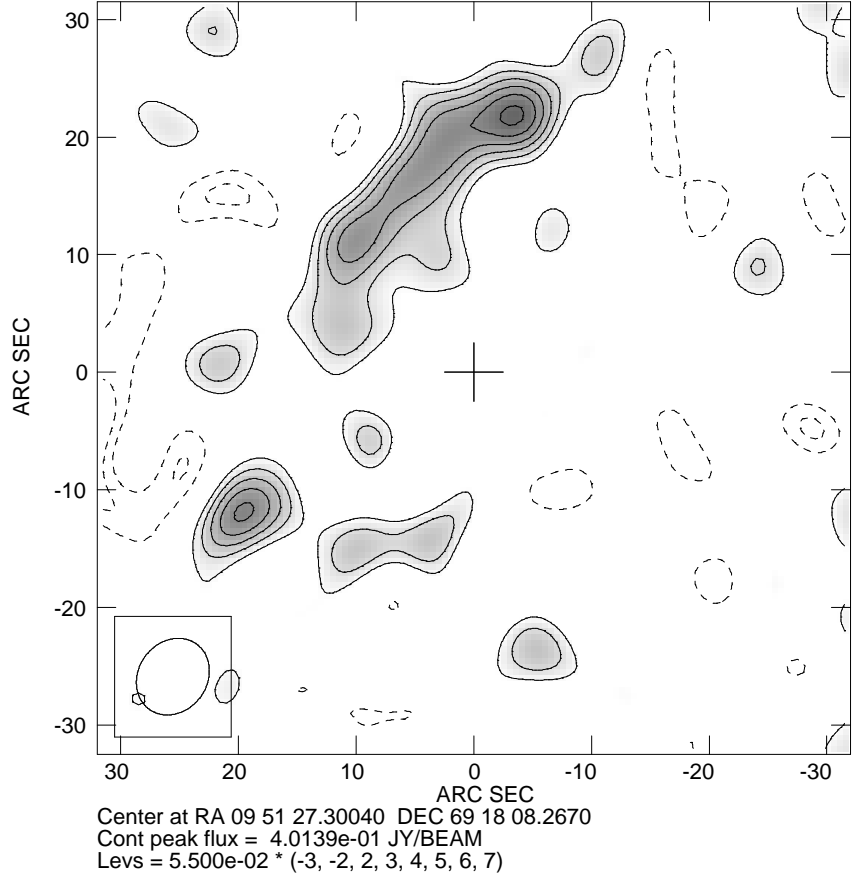}       
\hfill}
\caption{
Integrated intensity map of CO(J=1--0) emission in the central kpc of M81. 
Continuum emission from the nucleus at the position of the cross 
has been subtracted. 
Contours are at $-3,-2,2,3,4,5,6$, and 7 $\times$ 1.1 Jy \beam\ \kms\ ($= 1\sigma$).
The contour interval and peak integrated intensity correspond 
to molecular gas column density of 17 and 120 \Msolpc, respectively.
The synthesized beam of $6\farcs9 \times 5\farcs8$ in FWHM is shown 
in the bottom left corner.
\label{f.sqash} }
\end{figure}

\clearpage

\begin{figure}[t]
{\hfill                 
\epsscale{0.6}
\plotone{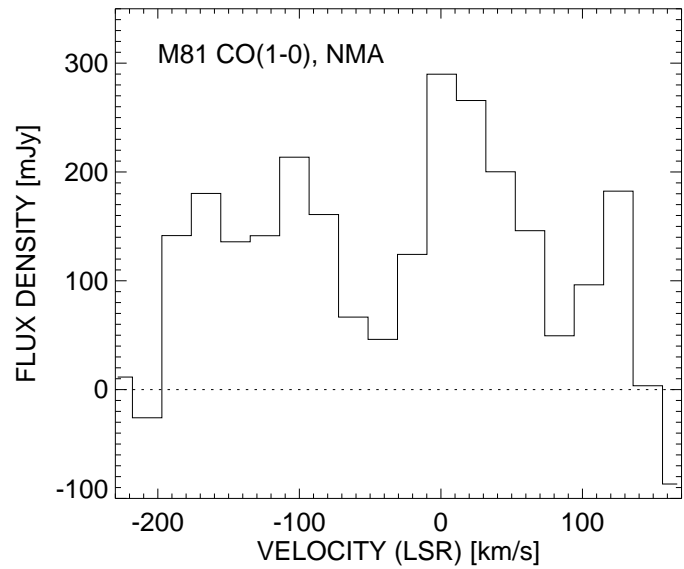}				
\hfill}
\caption{Spectrum of CO(J=1--0) emission in the central kpc of M81, observed in the
1\arcmin\ (FWHM) primary beam of the NMA.
The line width is $\sim 350$ \kms\ and the detected CO flux is
50 Jy \kms, which is about half of the single-dish flux observed 
at the NRAO 12 m telescope.
Continuum emission from the nucleus has been subtracted.
\label{f.COspec} }
\end{figure}

\clearpage

\begin{figure}[t]
{\hfill                 
\epsscale{1.0}
\plottwo{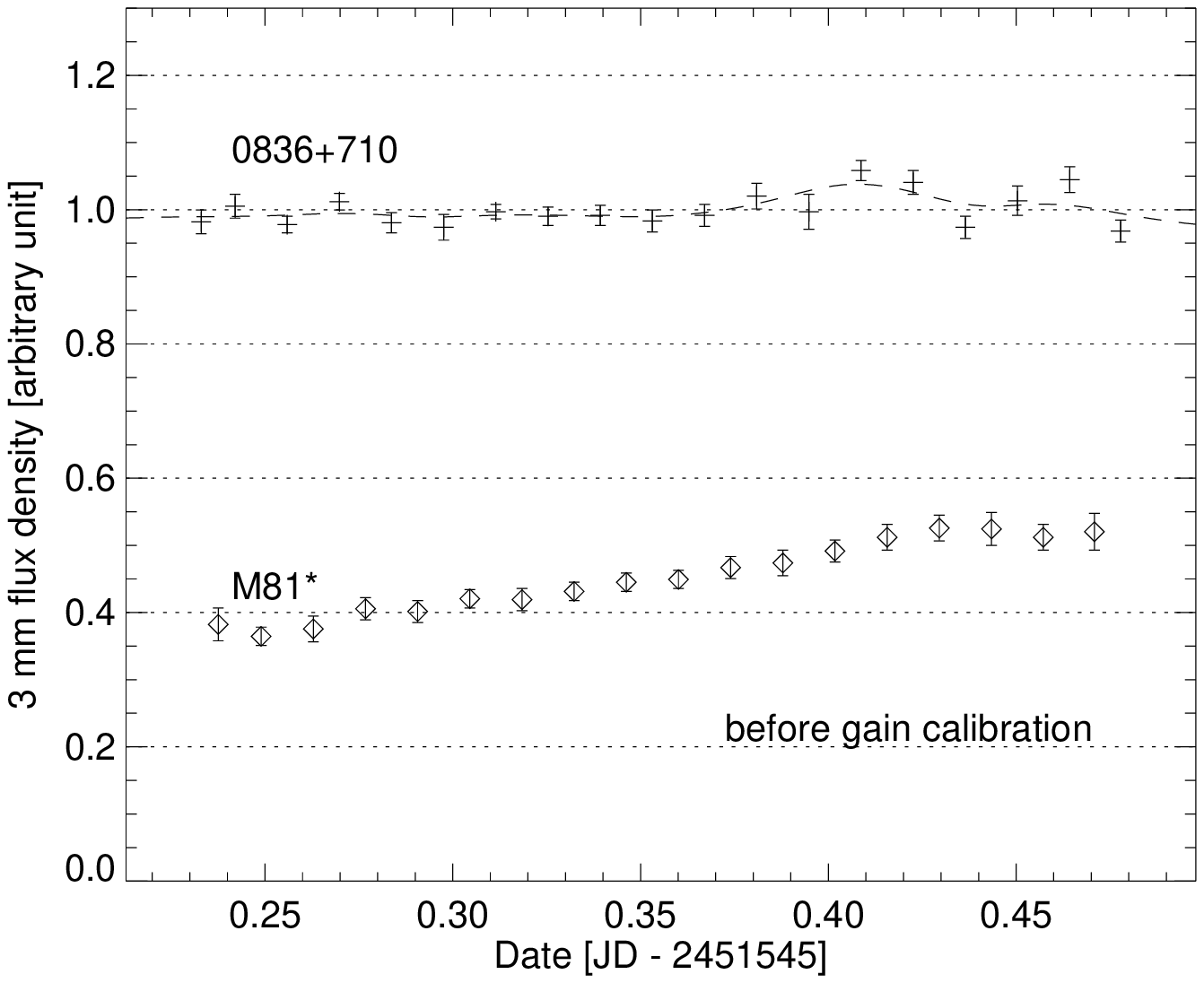}{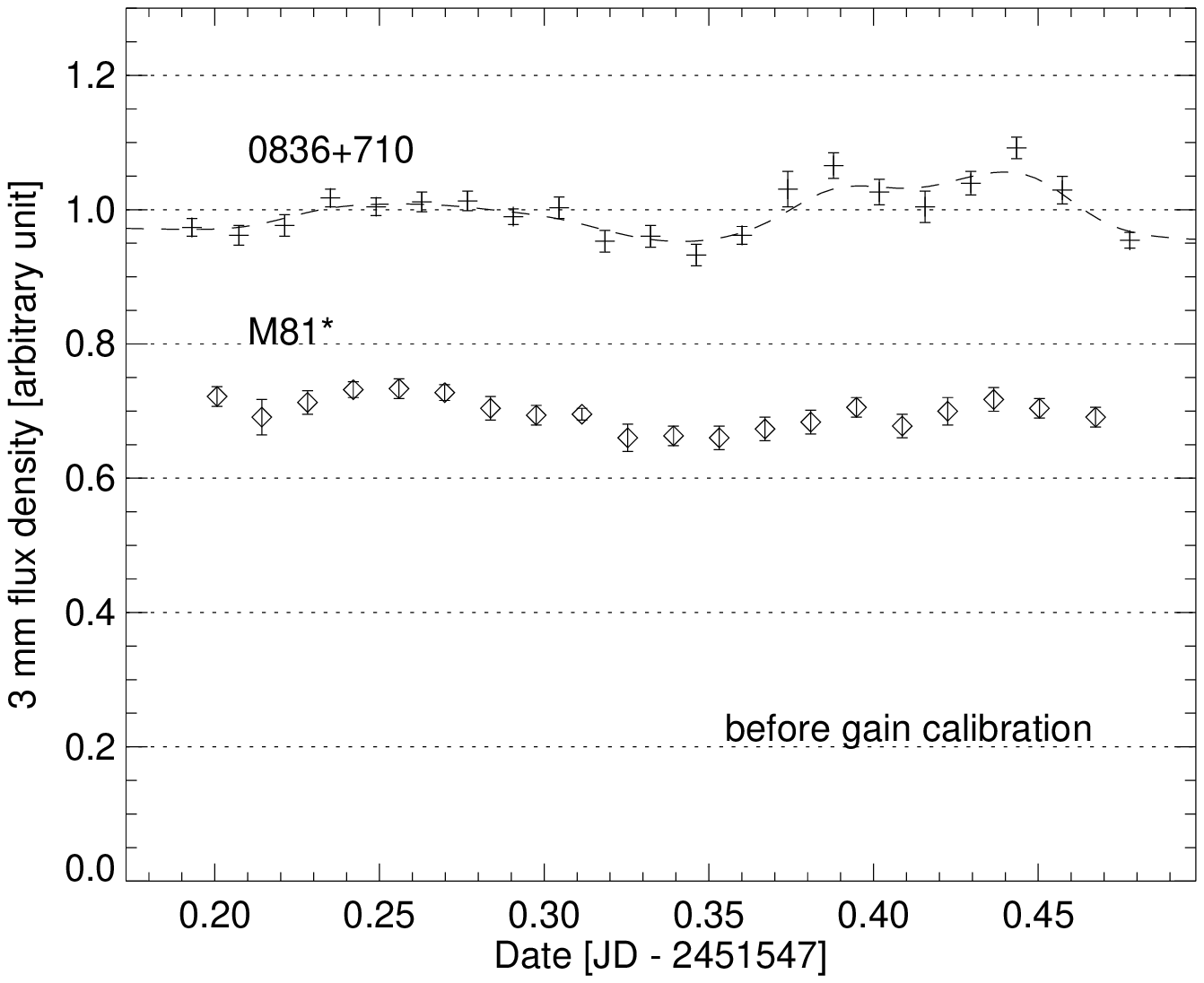}       		
\hfill}

{\hfill                 
\epsscale{1.0}
\plottwo{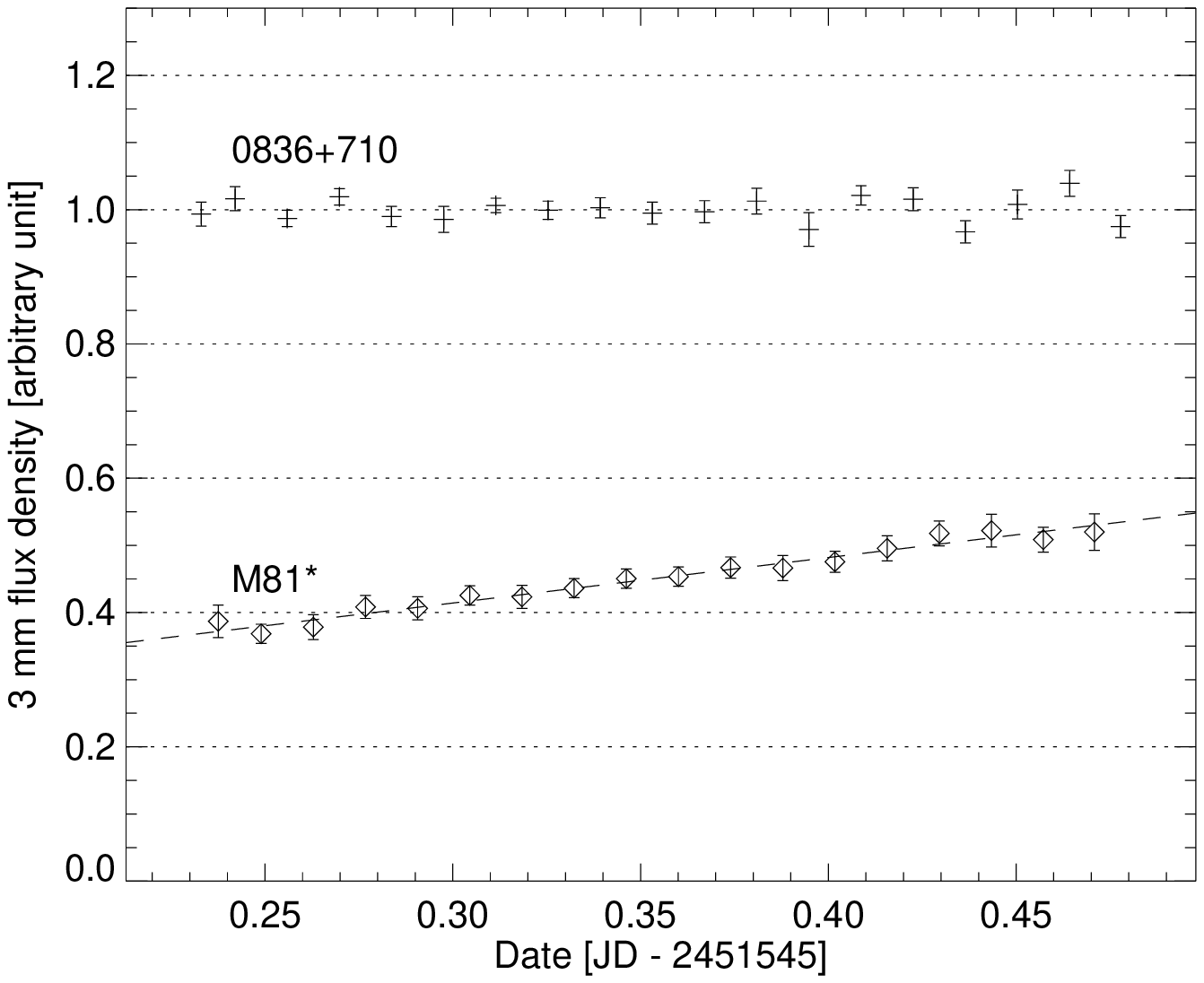}{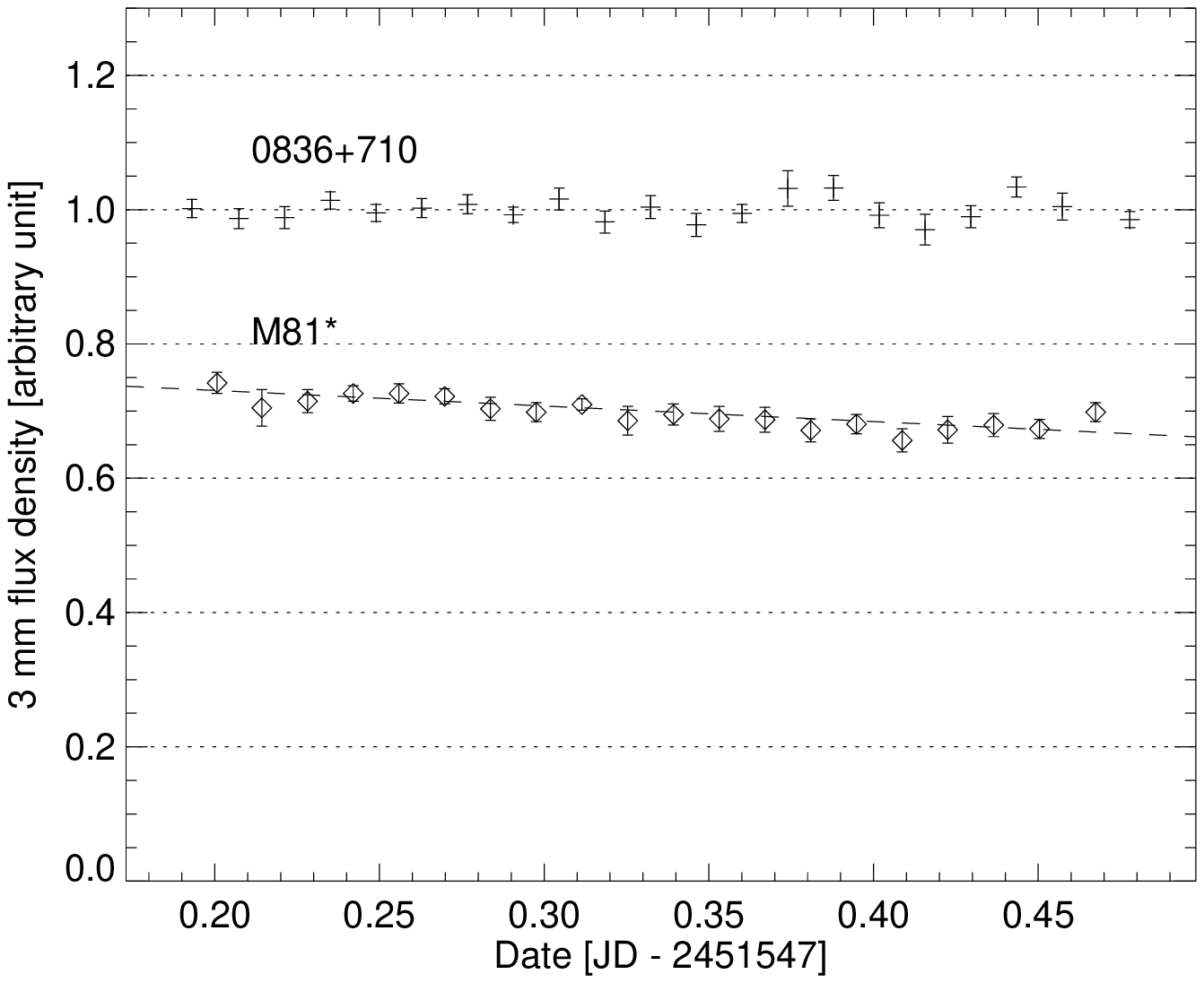}       	
\hfill}
\caption{
Time variation of the 3 mm continuum from \nuc\ on J.D.= 2451545 (left) 
and 2451547 (right). 
Top panels show data after chopper-wheel correction 
and before gain calibration.
Gain curves (shown as dotted lines) are derived 
by smoothing amplitudes of B\calib\ with a Gaussian kernel having  
$\sigma = 0.015$ day and a cutoff at $|\Delta t| = 0.050 $ day.
The large gain variation in the right panel, on J.D. = 2451547, 
is most likely due to coherence loss.
Bottom panels show data after applying the gain calibration.
A linear fit of \nuc\ is plotted over the calibrated data.
The monotonic variation of \nuc\ is clearly seen.
Error bars are $\pm 1 \sigma$ for \calib\ and $\pm 2 \sigma$ for \nuc. 
\label{f.D5cont} }
\end{figure}

\clearpage

\begin{figure}[t]
{\hfill                 
\epsscale{0.8}
\plotone{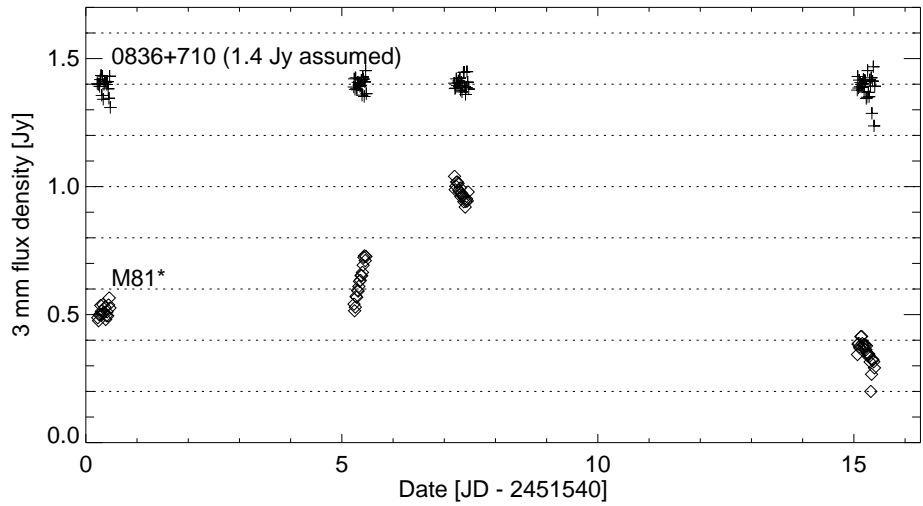}     
\hfill}
\caption{
Time variation of the 3 mm continuum from \nuc\ around the flare on 
J.D. $\sim 2451546$.
Instrumental gain has been calibrated assuming that \calib\ had 
a constant flux density of 1.4 Jy and was unpolarized.
\label{f.flare} }
\end{figure}

\clearpage

\begin{figure}[t]
{\hfill                 
\epsscale{0.7}
\plotone{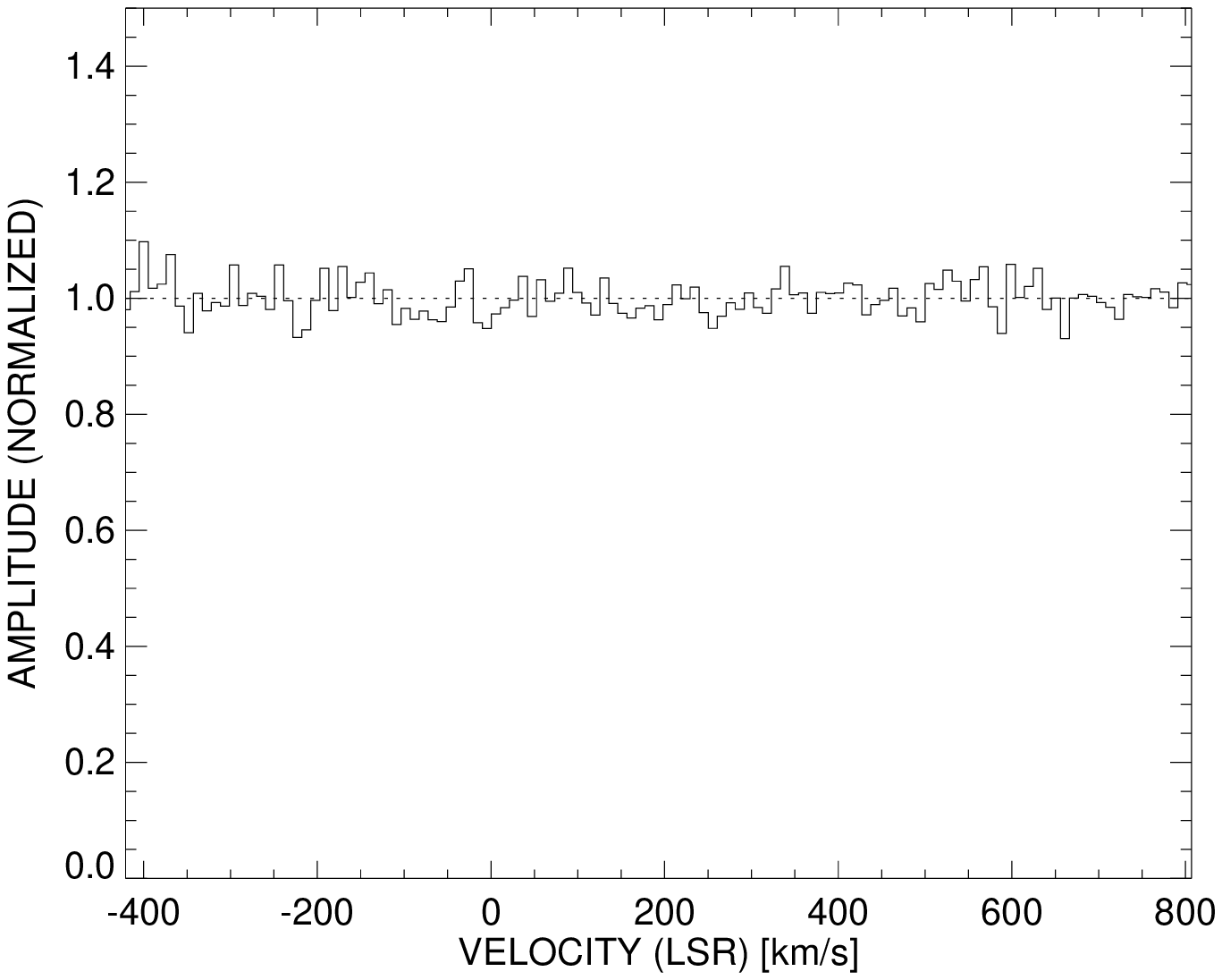}       		
\hfill}

{\hfill
\epsscale{0.7}
\plotone{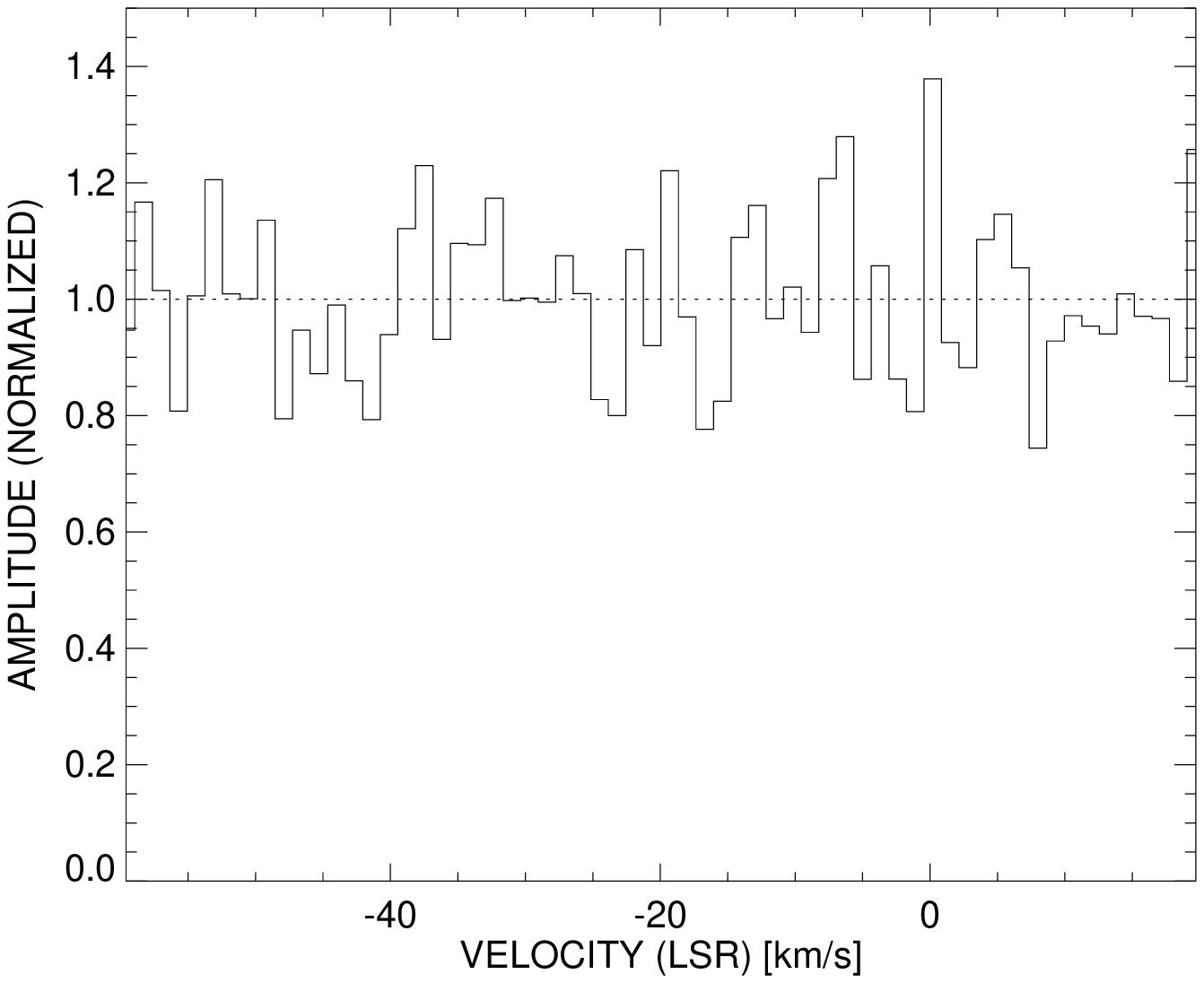}       		
\hfill}
\caption{
Spectra of the 3 mm continuum from M81$^{*}$ used to search 
for CO(J=0--1) absorption toward the nucleus.
The top panel has a 10.4 \kms\ resolution (obtained from UWBC), 
while the bottom panel has a 1.3 \kms\ resolution (observed with FX). 
A linear baseline is subtracted and amplitude is normalized to unity.
The systemic velocity of M81 is $-34$ \kms\ (LSR) 
and Galactic HI clouds have velocities around 3 \kms\ (LSR) 
in the direction of M81. No absorption is seen around these velocities.
The rms of the UWBC spectrum is 0.0315 and that of the FX spectrum is 0.140.
\label{f.contspec} }
\end{figure}

\clearpage

\begin{figure}[t]
{\hfill 
\epsscale{0.8}                
\plotone{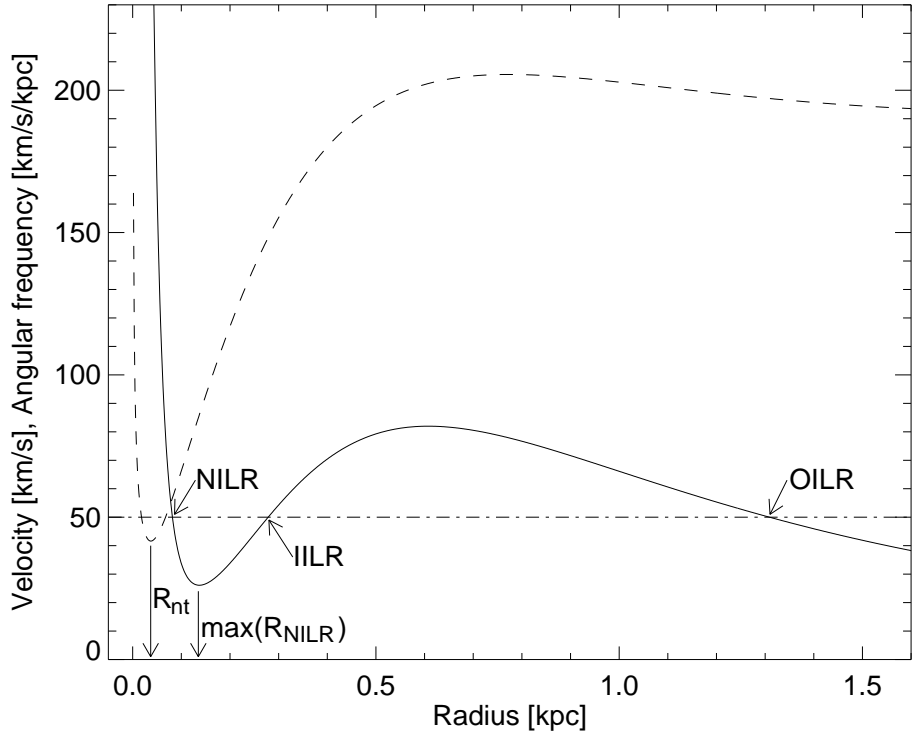}       
\hfill}
\caption{
The rotation curve (dashed line) and $\Omega - \kappa /2$ curve (solid line) 
for a model potential with a $10^7$ \Msol\ black hole at the galactic center.
The minimum of the rotation curve at \rk\ and the minimum of 
$\Omega - \kappa /2$ at $R = \max(R_{\rm NILR})$ are indicated.
Also marked are the nuclear inner Lindblad resonance (NILR), 
inner inner Lindblad resonance (IILR), 
and outer inner Lindblad resonance (OILR) 
for a bar pattern speed of 50 \kmskpc\ (dot-dashed line).
\label{f.rotcur} }
\end{figure}

\clearpage

\begin{figure}[t]
{\hfill 
\epsscale{1.0}                
\plotone{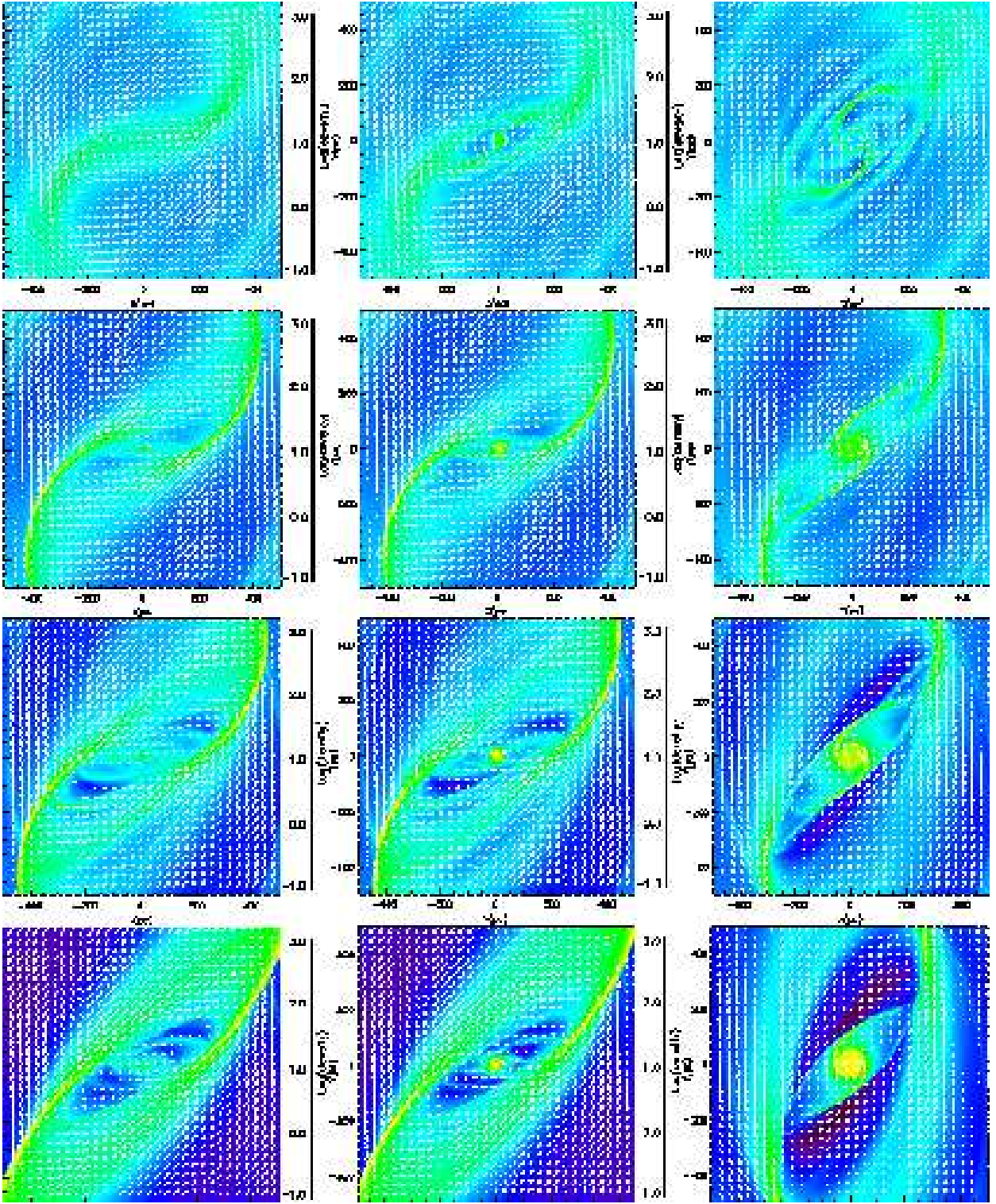}       
\hfill}
\caption{
Simulations of gas dynamics in the galactic center with a massive black hole 
and bar perturbations.
Each panel shows gas surface density in pseudocolor, 
and gas velocities with arrows, in the frame rotating with the bar.
The major axis of the bar is horizontal and the rotation of the galaxy 
is counterclockwise.
The three columns are, from right to left, for the mass of 
the central black hole of $10^6$, $10^7$, and $10^8$ \Msol\, respectively.
The four rows are, from the top, at $T = 20$, 40, 60, and 86 Myr.
The orbital period of the system at the radius of 500 pc is 15 Myr.
\label{f.sim} }
\end{figure}

\end{document}